\documentclass[conference]{IEEEtran}
\IEEEoverridecommandlockouts
\usepackage{cite}
\usepackage{amsmath,amssymb,amsfonts}
\usepackage{graphicx}
\usepackage{textcomp}
\usepackage{comment}
\usepackage[T1]{fontenc}
\usepackage[utf8]{inputenc}
\usepackage{dblfloatfix} 
\usepackage{algorithm}
\usepackage{algpseudocode}
\usepackage{soul}
\soulregister\mathcal{1}
\soulregister\subseteq{1}
\usepackage{xcolor}
\usepackage{amsmath,amssymb,booktabs,tabularx}
\usepackage[none]{hyphenat} 

\def\BibTeX{{\rm B\kern-.05em{\sc i\kern-.025em b}\kern-.08em
    T\kern-.1667em\lower.7ex\hbox{E}\kern-.125emX}} 
\begin{document}

\title{Resource Slicing for Time-Sensitive Data Traffic in Non-Terrestrial Networks with FSO ISLs
} 
\author{%
\IEEEauthorblockN{%
Abid Afridi,
Alexis A. Dowhuszko,
Risto Wichman,
Jyri H\"am\"al\"ainen}
\IEEEauthorblockA{Department of Information and Communications Engineering, Aalto University, Finland.}

\IEEEauthorblockA{Emails: \{abid.afridi, alexis.dowhuszko, risto.wichman, jyri.hamalainen\}@aalto.fi} \vspace{-10mm}

\thanks{This work was funded by the FOCAL, a doctoral network of the EU’s Horizon Europe programme in the frame of the MSCA GA no.~101169042.}
}
\maketitle
\begin{abstract}
In multi-hop Low Earth Orbit (LEO) satellite networks, time-sensitive packets experience queueing across multiple satellite hops, causing per-hop waiting times that accumulate and make the maximum end-to-end (E2E) delay difficult to guarantee. This paper presents a novel multi-hop slicing framework that reserves time-frequency resources for each time-sensitive flow of data originating at each satellite, aiming to maintain the reservation of communication resources along every inter-satellite link (ISL) on its route to the ground station (GS). The number of time-frequency resource blocks (RBs) reserved per-hop is determined from the data packet arrival statistics, E2E delay budget, and target level of reliability. Reservations on consecutive links are temporally aligned to enable pipelined forwarding of the time-sensitive data flows aggregated from terrestrial user terminals on the different LEO satellites. We evaluate the proposed framework against conventional scheduling and resource-allocation baselines, including pure time-division multiple access (Pure-TDMA), first-in-first-out (FIFO), and farthest-satellite-first (FSF), over the same multi-hop LEO satellite path. Under light traffic conditions, the baseline schemes provide comparable delay performance; however, their high-percentile delays increase sharply as data traffic load grows, whereas the proposed framework exhibits substantially slower delay growth. At the high arrival rate and high reliability target considered, the proposed scheme requires about 31~ms, a factor of 2.5–6.8 lower than the 76, 180, and 210~ms required by FIFO, FSF, and Pure-TDMA.
\end{abstract} 
\vspace{1mm}
 \vspace{-2mm}
\section{Introduction}
\label{sec:intro}
The standardization of non-terrestrial networks (NTN) in 3GPP Release 17 and the deployment of mega-constellations of Low Earth Orbit (LEO) satellites such as Starlink, OneWeb, and Kuiper have established LEO satellites as an integral component of 5G-Advanced and 6G systems. Their global coverage enables connectivity in regions where terrestrial infrastructure is unavailable, making them attractive for latency-sensitive services.
Although LEO satellites significantly reduce propagation delay compared with Geostationary Earth Orbit (GEO) satellite systems, the inability of all satellites to maintain continuous visibility of ground stations (GSs) in LEO constellations requires inter-satellite links (ISLs) with queues for data forwarding. Consequently, waiting time accumulates at each intermediate hop and often dominates the overall latency budget of the data traffic flows, especially under heavy traffic conditions~\cite{soret2020latency}. Measurements from
operational LEO satellite systems further demonstrate the importance 
of guaranteeing the end-to-end~(E2E) latency with high probability. For example, it was reported in \cite{mohan2024multifaceted} that the median of the minimum RTT of Starlink hovered around 40–50 ms in most countries, with regional variation (e.g., the 75th percentile exceeding 100 ms and tails reaching 200 ms in some under-provisioned regions, while remaining below 50 ms in well-provisioned regions). Under active download load, the measured RTT increased by roughly 2–4×, reaching almost 400–500 ms, based on 19.2 million M-Lab measurements from 34 countries. These observations
highlight that average (or median) E2E latency values alone are insufficient for
time-sensitive traffic, where the so-called right-hand side (upper) tail of its cumulative distribution function~(CDF) determines
the probability that a packet arrives within its required maximum delay constraint. For mission- and safety-critical NTN services, such as disaster monitoring, emergency response, and public-safety communications, stringent E2E latency and reliability are essential, since packet loss or arrival beyond the target maximum latency required by the upper-layer application may result in a critical failure of the whole process. These requirements motivate E2E resource coordination across the multi-hop mesh network that is configured on the sky among LEO satellites, in order to control queueing delay while satisfying the
latency and reliability requirements.

Most of the work reported so far in the literature has focused on routing and traffic control on the mesh satellite network to
reduce latency. The authors of \cite{han2023load} proposed a distributed back-pressure scheme
whose link weight is the remaining hop count from the next hop to the
destination, aimed at balancing the regional load to postpone congestion. We note that the algorithm proposed in [4] makes the forwarding decision per packet and
hop and, therefore, gives no guarantee of E2E latency at the flow level. Similarly, in \cite{ran2025graphpr}, the authors
encoded one-hop observations with a graph attention network so that each
satellite implicitly infers multi-hop queue and topology state before forwarding,
 while Zhang \textit{et al.} in \cite{zhang2025grlr} combined graph learning with
reinforcement learning for mega-constellations under time-varying ISL states. The approaches proposed in \cite{ran2025graphpr} and \cite{zhang2025grlr} lower the average delivery time and the packet loss
rate, but they aim at optimizing an aggregate expectation of the E2E delays of packets transported on the mesh network rather than ensuring a per-flow
guarantee. Finally, Wan \textit{et al.} routed deadline-constrained traffic over ISLs to GSs using a rolling-horizon approach, improving network throughput while satisfying delay requirements \cite{wan2021downlink}. However, their approach does not provide an E2E probabilistic latency guarantee. 

Scheduling and slicing studies, in turn, allocate resources
at individual links without coordinating them along the
multi-hop route \cite{afridi2026load, liao2025end, s25144327, liu2025reconfigurable, he2024digital, 8853247}. In~\cite{liao2025end}, the authors proposed an E2E cross-domain slicing
framework for LEO satellite Internet of Things that distributes delay budgets across
network domains and dynamically allocates resources using deep
reinforcement learning (DRL). Liu \textit{et al.}
developed an SDN-based framework for time-sensitive traffic in LEO
networks using differentiated forwarding, queue management, and
priority-aware scheduling~\cite{s25144327}. While these approaches
address time-sensitive traffic, they rely on
dynamic network-level scheduling rather than pre-allocating
flow-specific resources across successive satellite hops. Similarly,
DRL-based radio access network (RAN) slicing~\cite{liu2025reconfigurable} and
digital-twin-assisted slicing~\cite{he2024digital} adapt resource
provisioning to time-varying service demands, but operate primarily at
the RAN or network-resource level. Most closely related to our formulation, García-Morales \textit{et al.} partitioned
time-frequency resources among latency-sensitive slices in a 5G RAN and
dimensioned them for deterministic and non-deterministic traffic \cite{8853247},
yet the reservation covers a single access hop for Industry 4.0 and does not extend across intermediate hops.

In this paper, we propose an E2E flow-level resource
reservation framework for time-sensitive data traffic in multi-hop LEO
satellite networks. For each flow originating at satellite \(k\), the
framework establishes flow-specific time-frequency resource block (RB) reservations
across all links on its multi-hop forwarding path to the GS, including the
free-space optical (FSO) ISLs and the radio frequency (RF) feeder link (FL). The reservation
size is dimensioned according to the traffic arrival rate, the maximum E2E delay
budget, and the target level of reliability, while reservations on consecutive hops
are temporally aligned to support pipelined packet forwarding. Unlike
conventional hop-by-hop scheduling, which allocates resources
independently at individual links, the proposed framework coordinates
flow-specific resources across the complete forwarding path to control
queueing delay and satisfy the E2E latency-reliability requirements.

The rest of the paper is organized as follows. Section~\ref{sec:system_model}
describes the system model and formulates the problem. Section~\ref{sec:proposed_2d_slicing} presents the proposed 2D
slicing scheme and the reservation algorithm. Section~\ref{sec:results} reports
the simulation results, and Section~\ref{sec:conclusion} concludes the paper.
\section{System Model and Problem Formulation} 
\label{sec:system_model} 
\subsection{Network and Traffic Model}
\label{subsec:network_traffic}
\begin{figure}[t]
    \centering
    \includegraphics[width=\columnwidth]{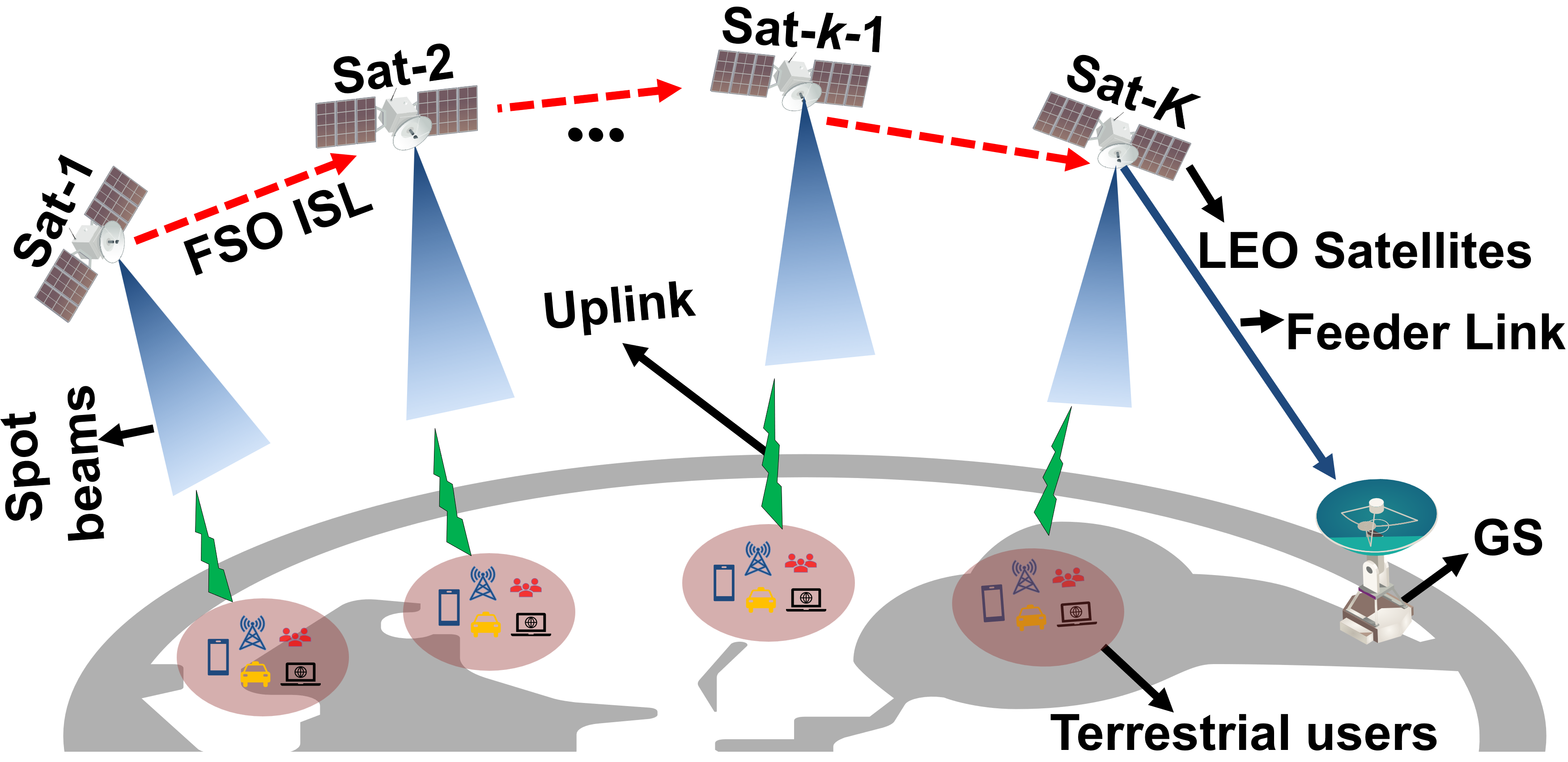}
    \caption{A multi-hop LEO satellite network, where each satellite \(k\) receives traffic from terrestrial users and forwards it over FSO ISLs toward satellite~\(K\), which delivers the aggregated traffic to the GS via the RF FL.}
    \label{fig:system_model} \vspace{-2mm}
\end{figure}

We consider a multi-hop forwarding path within a LEO satellite
constellation, as illustrated in Fig.~\ref{fig:system_model}.
The path contains satellites indexed by
\(k\in\mathcal{K}\) and a GS
\(g\), as the target destination of the considered
path. Each satellite aggregates time-sensitive traffic generated by the terrestrial terminals in its coverage area. The aggregate traffic entering the forwarding path at the satellite is referred to as flow~\(f\), where \(f\in\mathcal F=\{1,\ldots,K\}\). Hence, \(f\) identifies the source satellite of a flow, whereas \(k\) identifies the link of the satellite under consideration when forwarding time-sensitive data traffic towards the destination GS.

The physical links of the forwarding path are represented by
\(\mathcal{L}=\{\ell_1,\ldots,\ell_K\}\), where \(\ell_k\)
denotes the FSO ISL from
satellite \(k\) to satellite \(k+1\). The last link, \(\ell_K\), denotes
the FL from satellite \(K\) to GS \(g\) that is implemented over RF bands according to state-of-the-art satellite systems. The route followed
by the data flow originated in satellite $k$ is defined as \(\mathcal{L}_k
=
\left\{
\ell_k,\ell_{k+1},\ldots,\ell_K
\right\}\), which is a subset of $\mathcal{L}$ containing the indexes of physical links used by the
data flow of satellite~$k$ to reach the GS. The number of links crossed by the data flow of satellite $k$ is given by 
\(\left|\mathcal{L}_k\right|
=
K-k+1.\)

The data link layer frames of both ISLs and FL are divided into $N_{\rm TS}$ time slots of equal duration,
indexed by \(i\).
Without loss of generality, we assume that all links are synchronized in time and provide
\(N_{\mathrm{RB}}\) RBs in the frequency domain in each time slot. Data flows are composed of equal-sized data units of size $D$ bits that, without loss of generality, fit in one time-frequency RB of each ISL/FL frame.

\begin{figure}[t]
    \centering
    \includegraphics[width=\columnwidth]{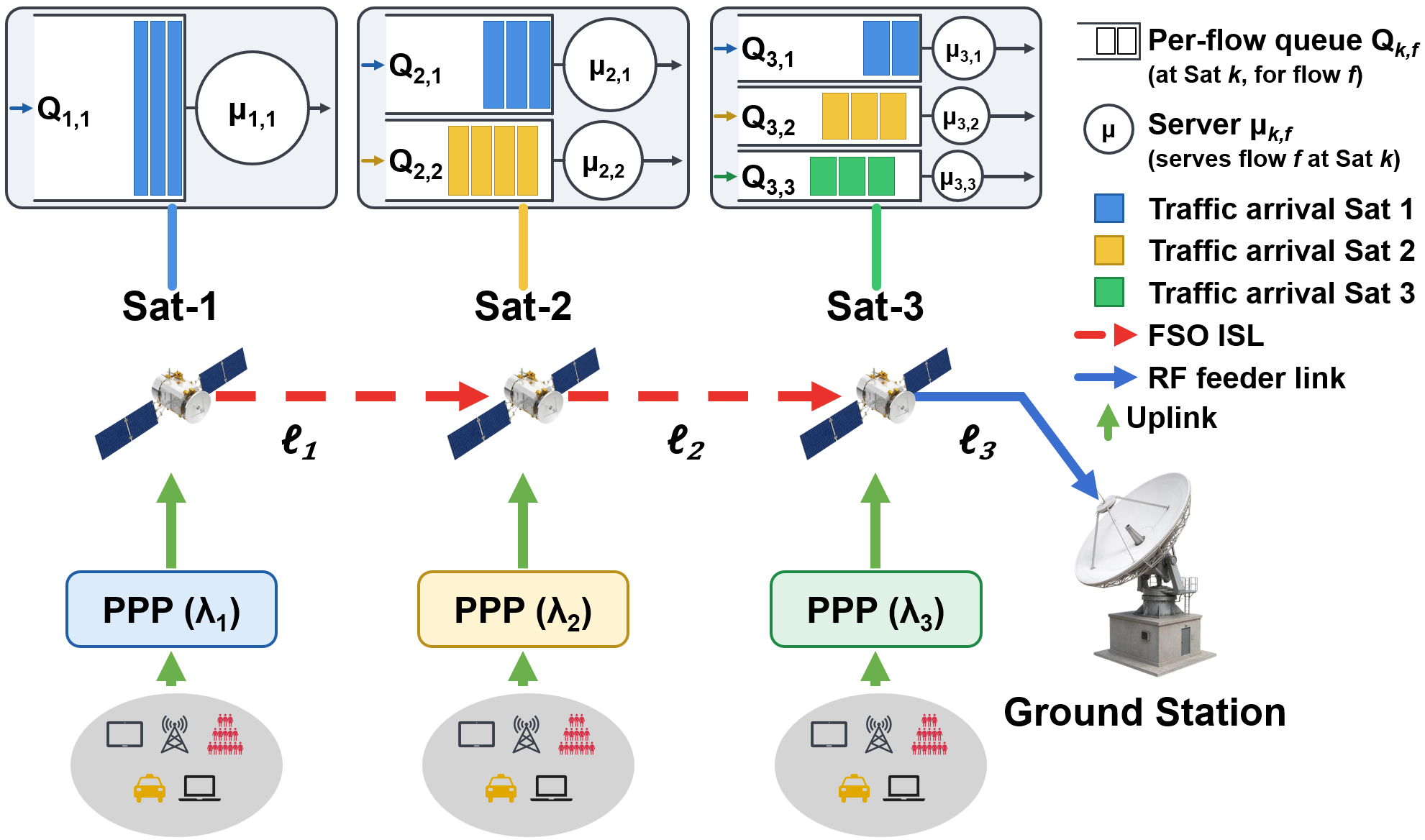}
    \caption{Three-satellite forwarding scenario, where each satellite \(k\) receives data traffic following a Poisson arrival process with mean \(\lambda_k\) from its associated terrestrial users. In this example, the multi-hop path comprises two FSO ISLs, \(\ell_1\) and \(\ell_2\), followed by the RF FL \(\ell_3\).} \vspace{-1mm}
    \label{fig:three_satellite_model}
\end{figure}

Each satellite aggregates traffic from ground user terminals, where the arrival of data packets is modeled as an independent Poisson process with mean rate $\lambda_k$. Based on this model, 
\begin{equation}
    A_k(i)
\sim
\operatorname{Poisson}
\left(
\lambda_{k}
\right),
\qquad
k\in\mathcal{K},
\label{eq:source_arrivals} \vspace{-2mm}
\end{equation}
can be interpreted as the number of packets generated by satellite
\(k\) at the beginning of time slot \(i\). We also assume that the locally generated arrival processes are independent.
Packets forwarded by the satellite with index $k$ during time slot \(i\) become available for transmission by
satellite \(k+1\) at the beginning of time slot \(i+1\). The traffic load
increases along the forwarding path because each satellite adds its
locally generated traffic to the packets received from upstream
satellites, as shown in Fig. \ref{fig:three_satellite_model}. 
\subsection{Delay Model}
\label{subsec:queue_delay}
We assume that each satellite has
sufficient buffer capacity such that the packet loss caused by queue overflow can be neglected. Hence, the actual
transmission of data flow $f$ available in satellite $k$ during time slot \(i\) can be expressed
\begin{equation}
u_{k,f}(i)
=
\min
\left\{
Q_{k,f}(i)+a_{k,f}(i),
\mu_{k,f}(i)
\right\},~\forall k\in \mathcal{K},
\label{eq:actual_transmission}
\end{equation}
where \(Q_{k,f}(i)\) is the queue backlog at the start of the time slot,
\(a_{k,f}(i)\) is the number of local arrivals originated at the same satellite (i.e., $f=k$) and arrivals of flows from the previous satellite (i.e., $f<k$), and \(\mu_{k,f}(i)\) is the
allocated number of RBs to serve the flow from satellite $k$ in time slot~$i$. Packets arriving at the beginning of time slot \(i\)
are eligible for transmission during the same slot. The queue backlog
is updated as \vspace{-2mm}
\begin{equation}
Q_{k,f}(i+1)
=
Q_{k,f}(i)
+
a_{k,f}(i)
-
u_{k,f}(i),~\forall k\in \mathcal{K}.
\label{eq:queue_evolution} 
\end{equation}
The delay experienced by a packet consists of two types: (i)~near-real time-varying delay, which represents the queueing delay, measured from the time the packet enters the queue of satellite
\(k\) until its transmission begins;
(ii) non-real-time (fixed) delay, which consists of processing delay, transmission delay,
and propagation delay and depends on the route geometry. Accordingly, the fixed delay associated with one link is defined as \vspace{-2mm}
\begin{equation}
\Delta\tau_{k}
=
\tau_{k}^{\mathrm{proc}}
+
\tau_{k}^{\mathrm{trans}}
+
\tau_{k}^{\rm prop},
\label{eq:fixed_link_delay}
\end{equation}
where \(\tau_{k}^{\mathrm{proc}}\) denotes the onboard processing delay required
to process a packet at the satellite, whereas \(\tau_{k}^{\mathrm{trans}}=D/B\) denotes the
packet transmission delay or time required to transmit the entire packet over the link, where \(D\) is the packet size
in bits, and \(B\) is the transmission rate of the link in bps. Finally,
\(\tau_{k}^{\mathrm{prop}}={L_\ell}/{c}\) denotes the propagation delay, where \(L_\ell\) is
the physical length of the link and \(c\) is the speed of light. These three components are assumed to introduce a fixed per-hop delay that the data flow needs to travel. The E2E delay that the packets of the data flow generated by satellite $k$
during time slot \(i\) experience until reaching the GS is given by
\begin{equation}
\tau_{k,f}^{\mathrm{E2E}}(i)
=
\sum_{j=k}^{K}
\left[
\Delta \tau_j + \tau_{j,f}^{\text{queue}}(i)
\right].
\label{eq:e2e_delay}
\end{equation}
Therefore, every data flow must reach the GS before the E2E deadline; hence, the latency-reliability requirement can be written as
\begin{equation}
\Pr
\left(
\tau_{k,f}^{\mathrm{E2E}}(i)>\tau_{\rm max}
\right)
\leq
\epsilon,
\qquad
k\in\mathcal{K},
\label{eq:latency_reliability}
\end{equation}
where \(\tau_{\rm max}\) is the maximum allowed E2E delay, \(\epsilon\) is
the maximum tolerated deadline-violation probability, and \(1-\epsilon\) is the
required latency reliability, which in this work can be interpreted as the probability that a data packet is delivered within the required time deadline. \vspace{-2mm}
\subsection{Reservation Model}
\label{subsec:reservation_model}
The reservation pattern is constructed over an allocation horizon of
$T_{\rm H}$ synchronized slots. Within this allocation horizon, the outgoing link of each satellite~$k$ uses a reservation window of the same length $W_{k}$ consecutive slots,
over which the reservation requirement on link~$\ell_k$ must be
satisfied. For a window starting
at time slot~$n$, the total number of RBs reserved for flow $f$ on link \(\ell_k\) is
defined as \vspace{-1mm}
\begin{equation} 
\bar{\mu}_{k,f}(n) = \sum_{i=n}^{n+W_k} \mu_{k,f}(i), \quad n= 0,\ldots, T_{\rm H} - W_k,
\label{eq:window_reservation} 
\end{equation}
where the starting time slot $n$ ensures that the complete window remains within the
allocation horizon. The total number of RBs reserved for the flows $f \in \mathcal{F}_k$ in satellite~$k$ during the given window starting at slot \(n\) is defined as \vspace{-2mm}
\begin{equation} 
\bar{\mu}_k(n) = \sum_{f \in \mathcal{F}_k}\bar{\mu}_{k,f}(n)
            = \sum_{f \in \mathcal{F}_k}  \sum_{i=n}^{n+W_{k}} \mu_{k,f}(i), 
\label{eq:total_window_reservation} 
\end{equation}
where $\mathcal{F}_k$ is the set of flows carried by link~$k$. Therefore, the actual window reservation must satisfy
\(\bar{\mu}_{k,f}(n) \geq~\bar{\mu}_{k,f}^{\min}\quad\forall k \in \mathcal{K}\),
where $\bar{\mu}_{k,f}^{\min}$ denotes the minimum number of RBs
required to serve data flow $f$ on link $\ell_k$
within every $W_k$-slot window. This service budget is selected
to limit the queueing delay so that the resulting E2E latency
satisfies the target reliability in \eqref{eq:latency_reliability}.
The actual reservation is distributed over the 2D time-frequency
grid while satisfying the per-slot RB capacity constraint. 
\subsection{Problem Formulation}
\label{subsec:problem_formulation}
We formulate a 2D reservation problem that provides flow isolation, hop-aware protection, and delay-aware coordination across consecutive links. The objective is to minimize the total number of RBs over the reservation window while meeting the E2E latency and reliability requirements of time-sensitive data traffic, and the optimization problem can be written as follows: \vspace{-2mm}
\begin{subequations}
\label{eq:problem}
\renewcommand{\theequation}{\theparentequation}
\begin{equation}
\underset{\boldsymbol{\mu}}
{\operatorname{min}}
\quad
\sum_{k \in \mathcal{K}}
{\bar{\mu}_k}(n)
\label{eq:reservation_objective}
\end{equation}
subject to:
\setcounter{equation}{0}
\renewcommand{\theequation}{\theparentequation\alph{equation}}
\begin{align}
&
\Pr\!\left(
\tau_{k,f}^{\mathrm{E2E}}(i)>\tau_{\rm max}
\right)
\leq
\epsilon, \qquad
\forall k\in\mathcal{K},
\label{eq:e2e_latency_reliability_constraint}
\\[1mm]
&
\sum_{i=n}^{n+W_k}
\mu_{k,f}(i)
\geq
\bar{\mu}_{k,f}^{\min},
\nonumber\\[-1mm]
&
\hspace{12mm}
\forall k\in\mathcal{K},
\quad
n=0,\ldots,T_{\rm H} - W_k,
\label{eq:window_reservation_constraint}
\\[1mm]
&
\sum_{f\in \mathcal{F}_k}\mu_{k,f}(i)
\leq
N_{\mathrm{RB}},
\quad
\mu_{k,f}(i)
\in
\mathbb{Z}^{+}, \quad \forall k\in\mathcal{K}.
\label{eq:link_resource_constraint}
\end{align}
\end{subequations}
In \eqref{eq:problem},
$\boldsymbol{\mu} =
\{\mu_{k,f}(i)\mid
k\in\mathcal{K},
f\in\mathcal{F}_k,
i\in\mathcal{N}\}$.
Constraint \eqref{eq:e2e_latency_reliability_constraint} represents the
E2E latency and reliability requirements. Constraint
\eqref{eq:window_reservation_constraint} ensures that each flow \(f\in\mathcal F_k\) is allocated at least \(\bar{\mu}_{k,f}^{\min}\) RBs on link \(\ell_k\) within every reservation window of \(W_k\) consecutive slots. Constraint
\eqref{eq:link_resource_constraint} ensures that the total per-slot reservation on each physical link does not exceed the available number of RBs \(N_{\mathrm{RB}}\) in any time slot, where \(F_k\) is the set of flows carried by link \(k\). \vspace{-1mm}

\section{Proposed 2D Slicing}
\label{sec:proposed_2d_slicing} 
\begin{figure}[t]
    \centering
    \includegraphics[width=\columnwidth]{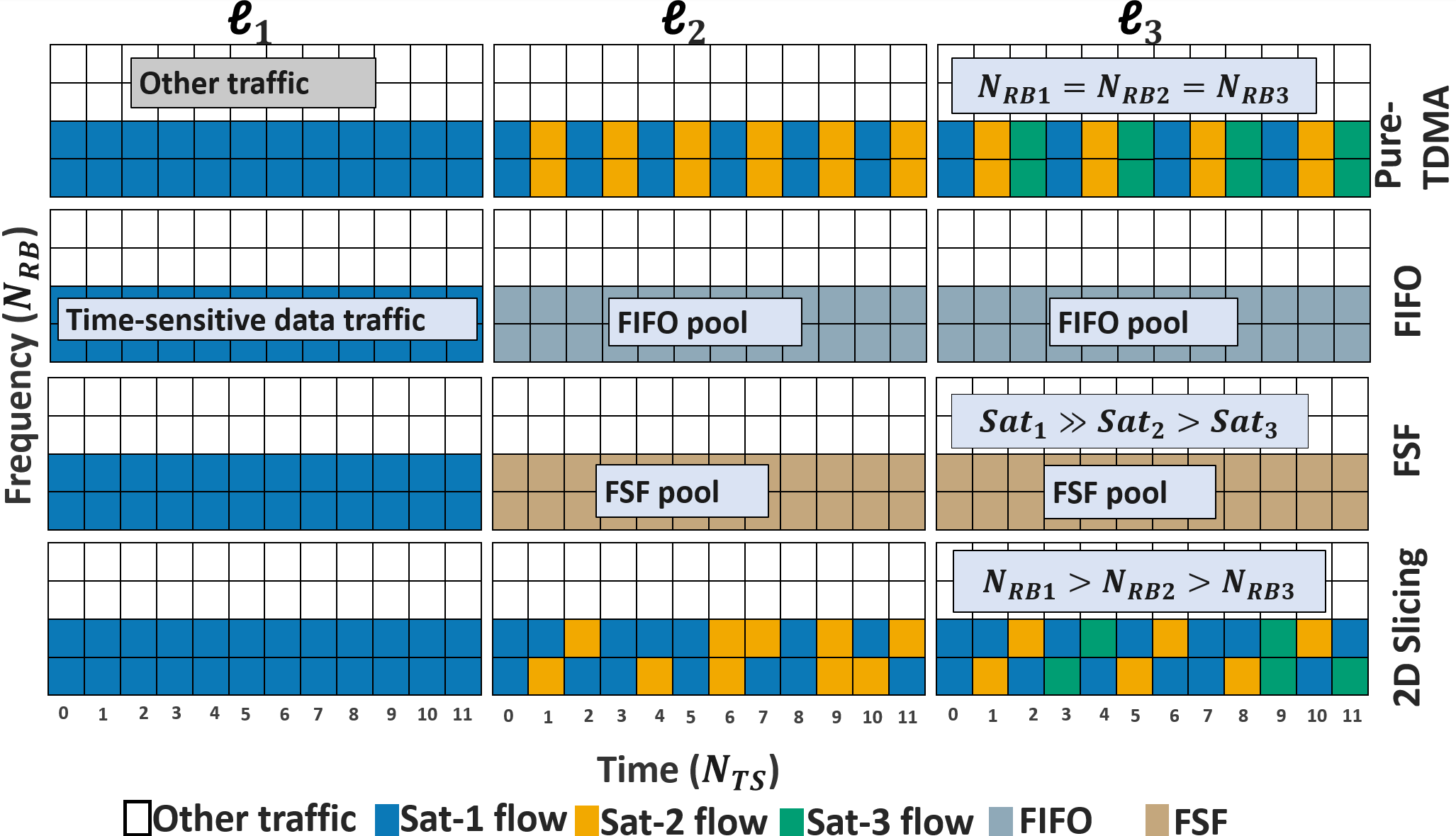}
    \caption{Time--frequency resource allocation across the forwarding links
    $\ell_{\rm 1}$, $\ell_{\rm 2}$ and $\ell_{\rm 3}$ for Pure-TDMA, FIFO, FSF, and the
    proposed 2D slicing. Each column corresponds to a link and each row to
    an allocation strategy.} \vspace{-1mm}
    \label{fig:SCHEMES}
\end{figure}

The proposed method partitions the time--frequency resource grid of each link into flow-specific allocations for time-sensitive data traffic, as illustrated in Fig.~\ref{fig:SCHEMES}. These 2D resource slices are pre-allocated along the forwarding path of each flow to provide isolation while satisfying its E2E targets. Upon arrival, a packet uses an available reserved RB for the data flow that corresponds to it; otherwise, it waits for the next reserved transmission opportunity, adding queueing delay. Under light data traffic, unused reserved RBs represent the resource overhead required to maintain the E2E performance.

Let \(\mathcal{B}_{\mathrm{U}}=\{1,\ldots,N_{\mathrm{U}}\}\) denote the time-sensitive data traffic RBs, where \(N_{\mathrm{U}}=\eta_{\mathrm{U}}N_{\mathrm{RB}}\). In the considered configuration, \(\eta_{\mathrm{U}}=0.5\) is the initial value, which means that half of the RBs of each FSO link are used by other kinds of traffic that are not time-sensitive. Unlike first-in-first-out~(FIFO) and farthest-satellite-first~(FSF) scheduling, which place all flows in a common resource pool, the proposed method assigns reserved RBs to individual time-sensitive flows. On a shared link, the method assigns a small additional margin to flows originating farther from the GS to reduce or eliminate their queueing delay. Thus, the reservation \(\bar{\mu}_k^*\) satisfies \vspace{-2mm}
\begin{equation}
\bar{\mu}_k^*\geq \bar{\mu}_{k}^{\min}, \qquad \bar{\mu}_{1}^*
\ge
\bar{\mu}_{2}^*
\ge
\cdots
\ge
\bar{\mu}_{K}^*, \qquad k\in\mathcal{K}.
\label{eq:2d_priority_order} \vspace{-1mm}
\end{equation}
Accordingly, Sat-1 receives slightly more protected service than Sat-2 on \(\ell_2\), while the reservation order on the FL \(\ell_3\) is Sat-1 \(>\) Sat-2 \(>\) Sat-3. Link \(\ell_1\) carries only the Sat-1 flow and, therefore, has an identical RB allocation under all scheduling methods. Reservation patterns on consecutive links are also aligned with the propagation and processing delays. This coordination enables pipelined forwarding, such that a packet transmitted over one link reaches the next queue close to its reserved service opportunity. The 2D resource reservation is summarized in Algorithm~\ref{alg:proposed}.

To compare the baselines, consider a shared link \(\ell_k\), which carries the
\(k\) flows with index $f \in \{1, \dots, k\}$ that have entered the path by satellite \(k\).
Let \(C_k\) denote the total number of RBs reserved for time-sensitive traffic on
\(\ell_k\). 
\begin{algorithm}[t]
\caption{End-to-End Flow-Level Resource Slicing}
\label{alg:proposed}
\begin{algorithmic}[1]
\Require $K$, $\mathcal{L}_k$, $N_{\mathrm{RB}}$, $\eta_{\mathrm{U}}$, $T_{\mathrm{H}}$,
$\lambda$, $\Delta\tau_k$, $\tau_{\max}$, $\epsilon$
\Ensure $\{\mu_{k,f}(i),\tau^{\mathrm{E2E}}_{k,f}\}$

\State Initialize $W_k$ and $\bar{\mu}_{k,f}^{\min}$ from
$(\lambda,|\mathcal{L}_k|,\tau_{\max},\epsilon)$, $\forall k$
\Repeat
    \State Reserve $\mu_{k,f}(i)$ over $W_k$ slots subject to
    (9b)--(9c)
    \State Align reservations on consecutive links by $\Delta\tau_k$
    \State Initialize $Q_{k,f}(1)\gets0,\ \forall k,f$
    \For{$i=1,\ldots,T_H$}
        \State Generate local arrivals $A_k(i)$ using (1)
        \For{$k=1,\ldots,K$}
            \State Compute $a_{k,f}(i)$ and $u_{k,f}(i)$ using (2)
            \State Update $Q_{k,f}(i+1)$ using (3)
        \EndFor
    \EndFor
    \State Evaluate $\tau^{\mathrm{E2E}}_{k,f}$ using (5)
\Until{all the flows satisfy (6)}
\State \Return $\{\mu_{k,f}(i),\tau^{\mathrm{E2E}}_{k,f}\}$
\end{algorithmic}
\end{algorithm}
Assume that all satellites generate traffic at the same rate, \(\lambda_k=\lambda\); hence, the aggregate load offered to link \(\ell_k\)
is \(k\lambda\).

Pure-TDMA divides the available RBs in link $\ell_k$ into \(k\) equal shares for all the flows, independently
of the demand. 
Thus, \(\ell_2\) gives one half to each of Sat-1 and Sat-2, and the FL \(\ell_3\) gives one third to each flow.
Each flow therefore
receives about \(C_k/k\) RBs, and its outage probability is given by \vspace{-2mm}
\begin{equation}
P_{\text{out},k}^{\text{TDMA}}(W_k) = 1- F_{\mathrm{Poisson}}
\left( \left\lfloor\frac{C_k(W_k)}{k}\right\rfloor;\ \lambda\, W_k \right).
\label{eq:tdma_shortage_probability}
\end{equation}
As the shares are fixed for each flow, the flows originating farther from the GS accumulate more queueing delay across the forwarding path. Also, the RBs left unused by one flow cannot be used by
another flow, so part of the reserved RBs may be wasted when the traffic load is light.

\begin{figure*}[t]
    \centering
    \resizebox{\textwidth}{0.37\textwidth}{%
        \includegraphics{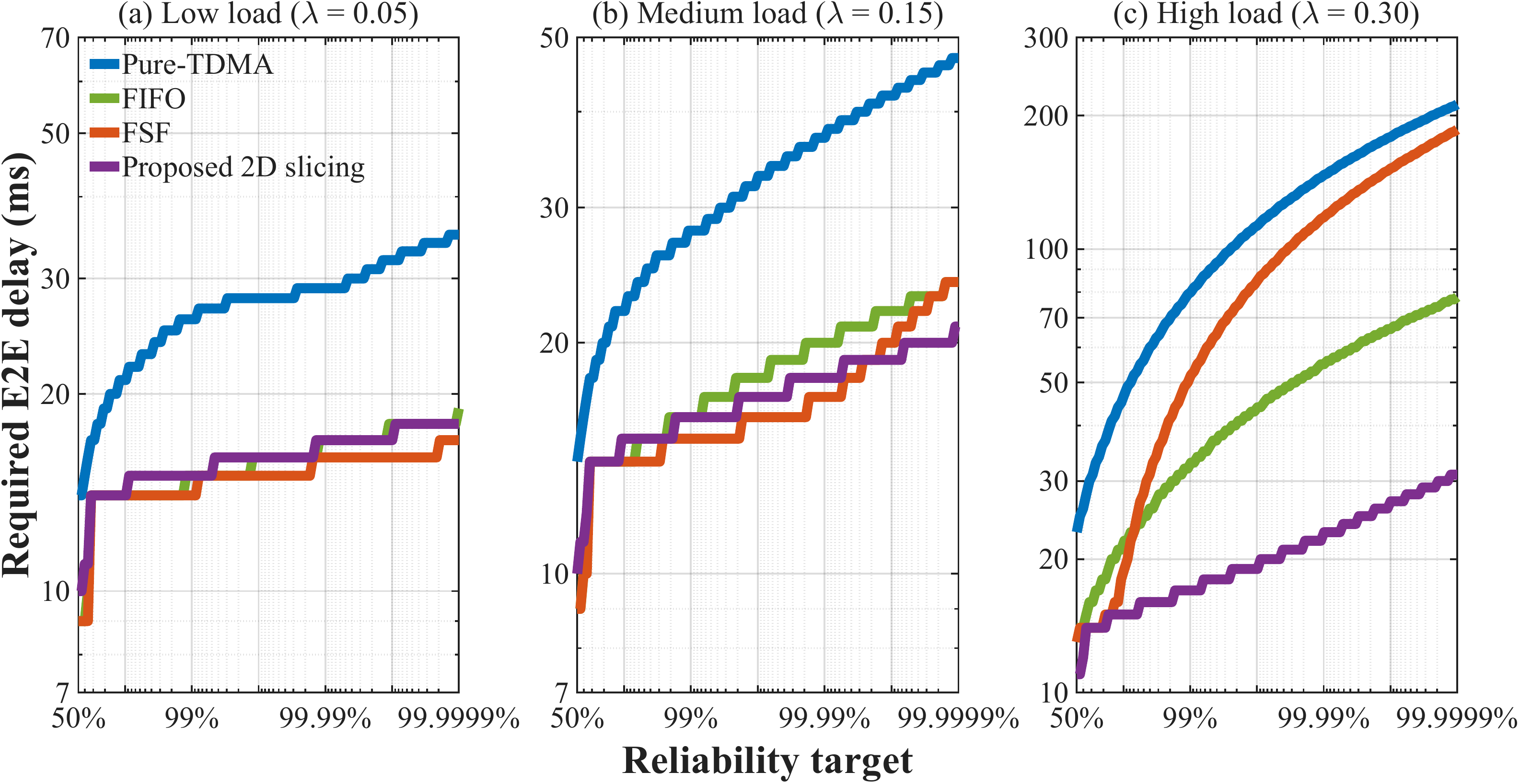}%
    }
    \caption{Required E2E delay as a function of the reliability target for
    Pure-TDMA, FIFO, FSF, and the proposed 2D slicing under
    (a) low (\(\lambda=0.05\)), (b)~medium (\(\lambda=0.15\)), and
    (c) high (\(\lambda=0.30\)) offered load.} 
    \label{fig:4} \vspace{-1mm}
\end{figure*}
FIFO places the \(k\) flows in one shared queue and serves them in order of
arrival. An overload condition caused by one flow will then affect the whole link, an event that will be common to every flow transported
on it. The outage probability will then be driven by the aggregate load \(k\lambda\): \vspace{-1mm}
\begin{equation}
P_{\text{out},k}^{\mathrm{FIFO}}(W_k) = 1- F_{\mathrm{Poisson}}
\left( C_k(W_k);\ k\lambda\, W_k \right).
\label{eq:fifo_overload_probability}
\end{equation}
As the offered load grows along the path, the FL \(\ell_K\) carries the largest
aggregate amount of data and is the first link to saturate. Under high load, its shared queue grows rapidly, which causes the long pooled tail delay on the FL.

FSF serves the flows on \(\ell_k\) in strict priority order, serving the farthest
satellite first. In the considered path, this places Sat-1 at the top of the order
and the newly entering flow at the bottom. The highest-priority flow
competes only against its own load \(\lambda\), so it is well protected with outage probability
\begin{equation}
P_{\text{out},k}^{\mathrm{FSF}}(W_k) = 1- F_{\mathrm{Poisson}}
\left( C_k(W_k);\ \lambda\, W_k \right), \quad {k=1}.
\label{eq:fsf_overload_probability}
\end{equation}
On the other hand, the lowest-priority flow is served only after all others, and
effectively faces the full aggregate data traffic with mean \(k\lambda\), so it is starved whenever the
link is busy. Strict priority therefore protects the upstream flows and pushes
the delay onto the flow closest to the GS, which in the three-satellite path is
Sat-3.\vspace{-1mm}
\section{Simulation Results}
\label{sec:results}
To evaluate the proposed approach for time-sensitive data traffic, we ran simulations in MATLAB using the Satellite Communications Toolbox, which provides tools to model and analyze satellite constellations, the GS, and communication links. In this simulation setup, we considered the LEO satellite forwarding path at an altitude of 550~km. We considered three satellites (Sat-1, Sat-2, and Sat-3) and one GS, as shown in Fig.~\ref{fig:three_satellite_model}. For each configuration, $1 \times 10^4$ independent Monte Carlo realizations were
performed, each with a warm-up period of $3\times10^3$ time slots to estimate the reported tail
statistics. Table~\ref{tab:simulation_parameters} summarizes the simulation parameters.

Fig.~\ref{fig:4} shows the required E2E delay versus the reliability
target for three offered loads, with a time-slot duration of $1$\,ms. The best-performing scheme depends on both
quantities. At low load (Fig.~\ref{fig:4}(a)), resource slicing provides no gain. FSF and FIFO achieve the lowest bounds; the three curves remain within about
$2$\,ms at the strict tail, and only Pure-TDMA requires a noticeably
larger bound, since the reserved RBs are underutilized at low arrival rates.
This advantage is retained only for loose targets as the load increases. In
Fig.~\ref{fig:4}(b), the curves intersect, and at $99.9999\%$, the proposed scheme
requires about $20$\,ms, compared with $24$\,{ms} for FIFO and
FSF and $49\,\mathrm{ms}$ for Pure-TDMA. Under high load (Fig.~\ref{fig:4}(c)),
it attains the lowest bound at every target, requiring about $31$\,{ms}
at $99.9999\%$, whereas FIFO, FSF and Pure-TDMA require $76$, $180$ and
$210$\,{ms}, respectively.

\begin{figure}[t]
  \centering
  \includegraphics[width=\columnwidth]{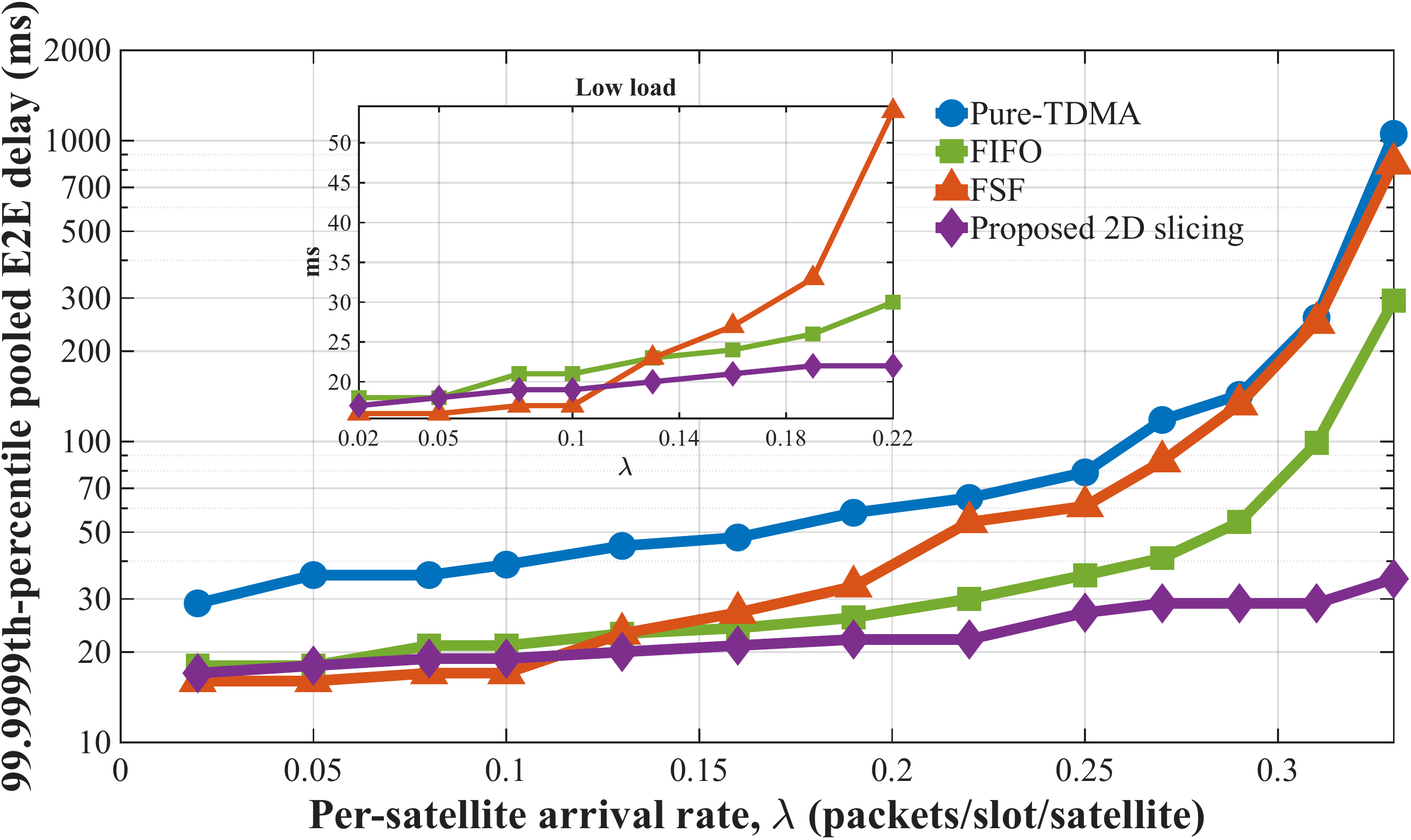}
  \caption{High-reliability (\(99.9999\)th-percentile) E2E delay with respect to the
per-satellite arrival rate \(\lambda\) for different schemes.} 
  \label{fig:5} \vspace{-2mm}
\end{figure}

Fig.~\ref{fig:5} plots the $99.9999$th-percentile E2E delay against the per-satellite Poisson arrival rate $\lambda$. The shared-queue baselines are marginally better only while the queues are short, and the curves
intersect near $\lambda=0.12$, beyond which the tail delay of the baselines
increases rapidly. This is explained by the shared queue of FIFO, the
starvation of the low-priority flow under FSF, and the fixed shares of Pure-TDMA, which approach saturation. The tail delay under FSF increases fastest and exceeds FIFO near
$\lambda=0.20$. The tail delay of the proposed scheme rises slowly, from about $18$ to
$35$\,{ms} over the considered arrival-rate range. 
At \(\lambda=0.33\), the proposed scheme requires approximately 35 ms, corresponding to delay reductions of approximately $88\%$, $96\%$, and $97\%$ relative to FIFO, FSF, and Pure-TDMA, respectively.

\begin{figure}[t]
    \centering
    \includegraphics[width=\columnwidth]{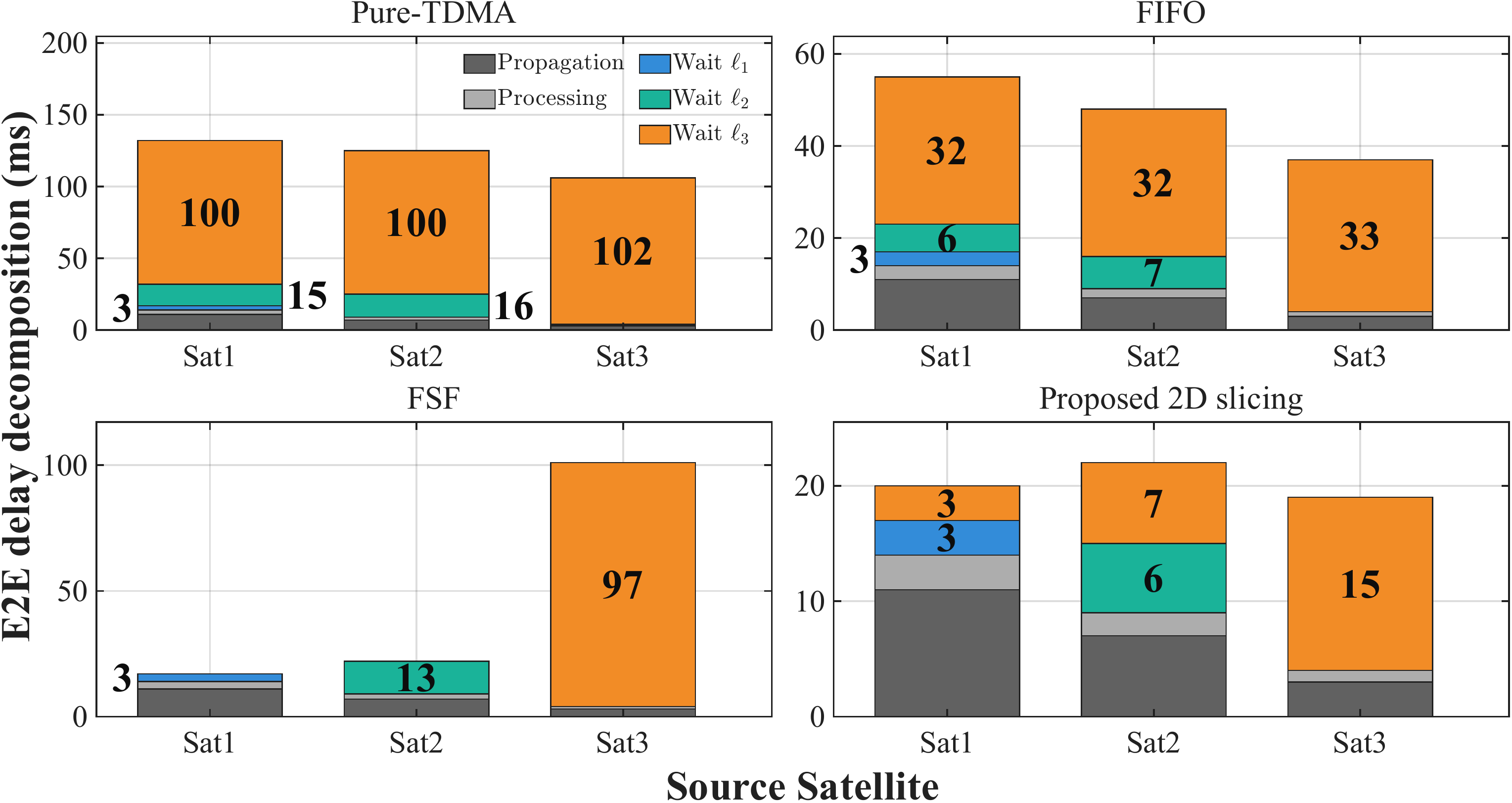}
    \caption{E2E delay decomposition for the three source satellites under
    Pure-TDMA, FIFO, FSF, and the proposed 2D slicing, showing propagation,
    processing, and per-link queueing delays along \(\ell_1\), \(\ell_2\), and
    \(\ell_3\).} 
    \label{fig:6} 
\end{figure}

Fig.~\ref{fig:6} decomposes the E2E delay of each source satellite into its
per-hop components. Propagation and processing are independent of the scheduling
scheme and decrease from Sat-1 to Sat-3, since a flow originating farther from the GS is relayed over more links. All schemes yield the same waiting time of about
$3\,\mathrm{ms}$ on $\ell_1$, which carries Sat-1 data flows alone; however, they differ only on
$\ell_2$ and $\ell_3$, where several flows are combined, and the effect is
most pronounced on the FL. Pure-TDMA incurs about $100$\,{ms} of waiting
time in the FL for every flow, as its fixed shares cannot follow the varying demand, and
FIFO reduces this to about $32$--$33$\,{ms} by pooling the feeder
capacity. FSF nearly eliminates the FL waiting of the higher-priority flows at
the expense of Sat-3, which waits approximately $97$\,{ms} on average. By reserving RBs in
the time--frequency grid for each flow, the proposed scheme limits the FL
waiting to about $3$, $7$ and $15$\,{ms}, so that the totals remain
within $19$--$22$\,{ms}. The FL queue is therefore the dominant
bottleneck of the multi-hop path.
\begin{figure}[t]
  \centering
  \includegraphics[width=\columnwidth]{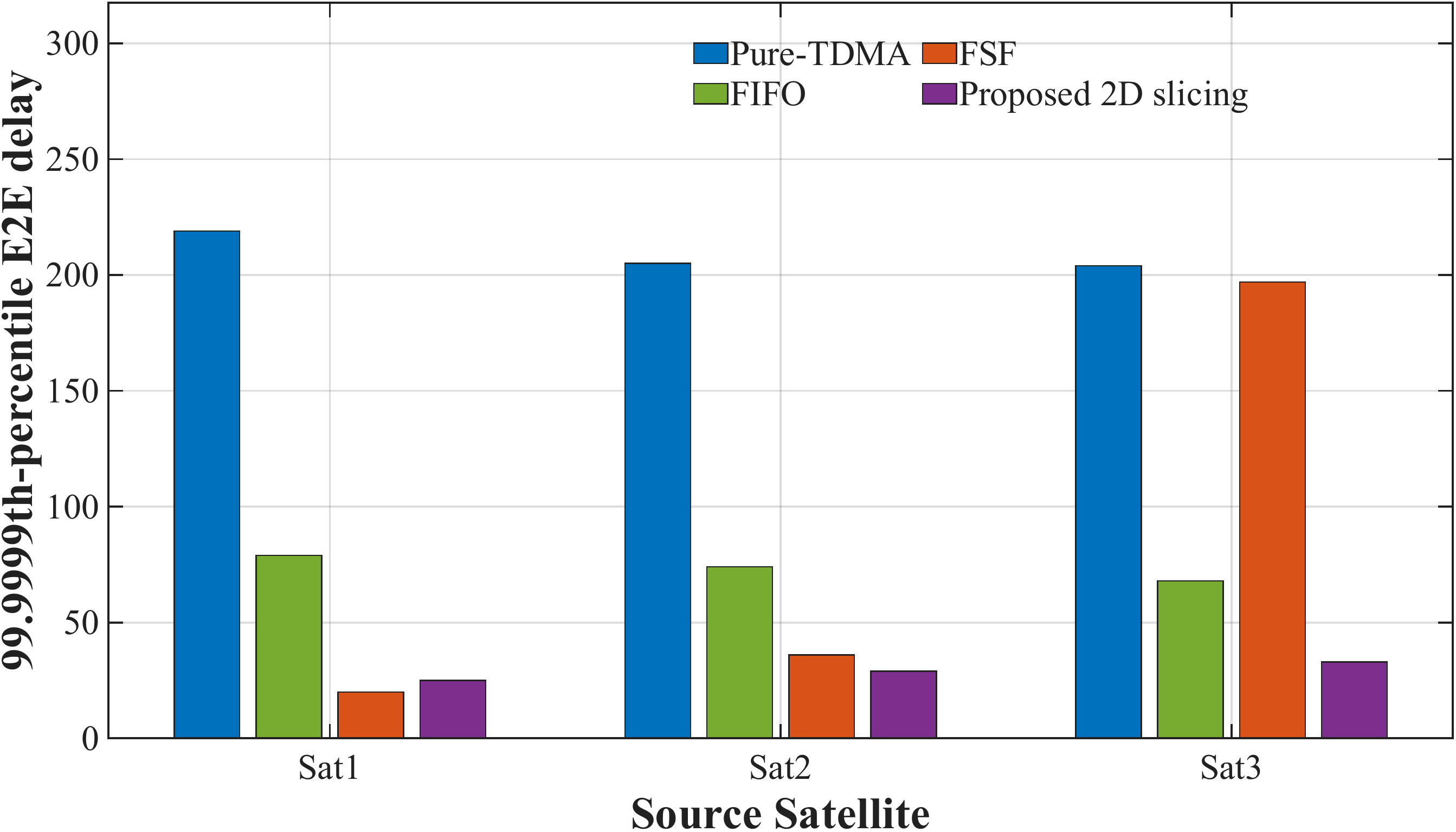}
  \caption{Comparison of the \(99.9999\)th-percentile E2E delay (ms) for traffic
originating at Sat-1, Sat-2, and Sat-3 under Pure-TDMA, FIFO, FSF, and the
proposed 2D slicing.} 
  \label{fig:7} 
\end{figure}

Fig.~\ref{fig:7} compares the $99.9999$th-percentile E2E delay of the three
source flows. Under Pure-TDMA and FIFO, the percentile decreases only slightly
from Sat-1 to Sat-3, since traffic from the later sources is relayed over fewer ISLs, but the tail
delay remains high, at about $204$--$218$\,{ms} and
$68$--$79$\,{ms}. FSF gives strict priority to flows originating farther from the GS,
reducing the Sat-1 tail delay to about $20\,\mathrm{ms}$ while that of Sat-3
increases to about $197$\,{ms}, a direct consequence of the starvation
experienced by the low-priority flow. The proposed scheme instead confines the
three tail delays to about $25$--$33$\,{ms}, reducing the Sat-3 tail delay by about $52\%$, $83\%$ and $84\%$ relative to FIFO, FSF and Pure-TDMA. It
therefore provides a low tail delay together with nearly uniform service across
paths of different lengths.  
\section{Conclusion and Future Work}
\label{sec:conclusion}
This paper presented an E2E resource slicing
framework for time-sensitive traffic in multi-hop LEO satellite
networks. By coordinating time-frequency reservations across the
forwarding path, the proposed method limits queueing delay and
provides reliable E2E latency performance. Simulation results
showed that the proposed scheme substantially outperforms Pure-TDMA,
FIFO, and FSF under high traffic loads, while remaining competitive
under light load. Future work will consider time-varying topology,
link-dependent capacities, mixed traffic classes, and DRL-based adaptive multi-hop resource allocation.

\begin{table}[t]
\caption{Simulation Parameters}
\label{tab:simulation_parameters}
\centering
\footnotesize
\renewcommand{\arraystretch}{1.1}
\setlength{\tabcolsep}{6pt}
\begin{tabular}{|p{0.46\linewidth}|p{0.12\linewidth}|p{0.28\linewidth}|}
\hline
\textbf{Parameter} & \textbf{Symbol} & \textbf{Value} \\
\hline
Satellite altitude & $h$ & 550 km \\
\hline
Number of satellites & $K$ & 3 \\
\hline
Number of ground stations & $G$ & 1 \\
\hline
Inter-satellite link type & -- & FSO \\
\hline
ISL distance & $L_{\mathrm{ISL}}$ & 1200 km \\
\hline
Nominal arrival rate & $\lambda$ & 0.02--0.33 pkt/slot/sat\\
\hline
Normalized packet demand & -- & 1 RB/pkt \\
\hline
Nominal warm-up period & -- & $3\times10^{3}$ slots \\
\hline
Nominal observation period & -- & $2\times10^{6}$ slots \\
\hline
Nominal Monte Carlo realizations & -- & $10^{4}$ \\
\hline
Reliability-target range & -- & 50--99.9999\% \\
\hline
\end{tabular} 
\end{table}
\bibliographystyle{IEEEtran} 
\bibliography{References}

\end{document}